\documentclass[conference]{IEEEtran}

\usepackage[ruled]{algorithm2e}
\usepackage{stmaryrd}
\usepackage{graphicx}
\usepackage{color}
\usepackage{comment}
\usepackage{url}
\usepackage{wrapfig}
\usepackage{subfigure}
\usepackage{amsmath}
\usepackage{enumitem}
\setlist[enumerate]{leftmargin=.5cm}
\setlist[itemize]{leftmargin=.3cm}

\newcommand{\superscript}[1]{\ensuremath{^{\textrm{#1}}}}

\def\wu{\superscript{*}}
\def\wg{\superscript{\dag}}

\begin{document}

\newcommand{\sysname}{ModelSink}
\title{Physiology-Aware Rural Ambulance Routing}

\author{\IEEEauthorblockN{Mohammad Hosseini\wu, Richard B. Berlin Jr.\wu\wg, Lui Sha\wu}
\IEEEauthorblockA{
  \begin{tabular}{ccc}
    \wu Department of Computer Science  & & \wg Department of Surgery \\
    University of Illinois at Urbana-Champaign  & & Carle Foundation Hospital \\
    \{shossen2, lrs\}@illinois.edu  & & richard.berlin@carle.com \\
  \end{tabular}
  ~\\
}}

\maketitle

\date{April 2017}
\maketitle

\begin{abstract}
The ultimate objective of medical cyber-physical systems is to enhance the effectiveness of patient care. During emergency patient transport from rural medical facility to center tertiary hospital, real-time monitoring of the patient in the ambulance by a physician expert at the tertiary center is crucial. The physician experts can provide vital assistance to the ambulance EMT personnel. While telemetry healthcare services using mobile networks may enable remote real-time monitoring of transported patients, physiologic measures and tracking are at least as important and requires the existence of high-fidelity communication coverage. However, the wireless networks along the roads especially in rural areas can range from 4G to low-speed 2G, some parts with communication breakage. From a patient care perspective, transport during critical illness can make route selection patient state dependent. Prompt decisions with the relative advantage of a longer more secure bandwidth route versus a shorter, more rapid transport route but with less secure bandwidth must be made. The trade-off between route selection and the quality of wireless communication is an important optimization problem which unfortunately has remained unaddressed by prior work.

In this paper, we propose a novel physiology-aware route scheduling approach for emergency ambulance transport of rural patients with acute, high risk diseases in need of continuous remote monitoring. We mathematically model the problem into an NP-hard graph theory problem, and approximate a solution based on a trade-off between communication coverage and shortest path. We profile communication along two major routes in a large rural hospital settings in Illinois, and use the traces to manifest the concept. Further, we design our algorithms and run preliminary experiments for scalability analysis. We believe that our scheduling techniques can become a compelling aid that enables an always-connected remote monitoring system in emergency patient transfer scenarios aimed to prevent morbidity and mortality with early diagnosis and effective treatment.

\end{abstract}
\IEEEpeerreviewmaketitle




\section{introduction}
There is a great difference in the provision of emergency medical care between rural and urban areas. The highest death rates are found in rural counties \cite{joms2016}. The lack of tertiary medical expertise and pre-hospital services in remote rural areas has motivated significant research effort in recent years to enhance the effectiveness of rural patient care.

Patient monitoring during emergency ambulance transport from rural areas to center tertiary hospitals requires reliable and real-time communication. This allows the physician experts in the center hospital to remotely supervise the patient in the ambulance and assist the Emergency Medical Technicians (EMT) to follow best treatment practices based on patient's physiological state. Unfortunately, remote monitoring of patients with critical illness involves real-time transmission of vital signs, graphs, and medical data, and relies on the presense of mobile network with high-fidelity communication. However, the wireless networks along the roads in rural areas can range irregularly from 4G to low speed 2G links, including some routes with cellular network communication breakage and only low speed satellite links \cite{satellite}. Due to the lack of access to satellite links, we do not address them in this paper. However, the technology presented in this paper can be easily extended to address satellite link. This poses a significant challenge for real-time monitoring of patients in an ambulance, therefore compromising the safety of patient care. The problem is exacerbated particularly in the high mobility scenarios of high-speed ambulances. Therefore, ensuring proper network Quality of Service (QoS) for the life-critical and communication-sensitive remote care especially for patients with acute diseases becomes crucial.

\begin{figure}[!t]
\centering
\includegraphics[width=1\columnwidth]{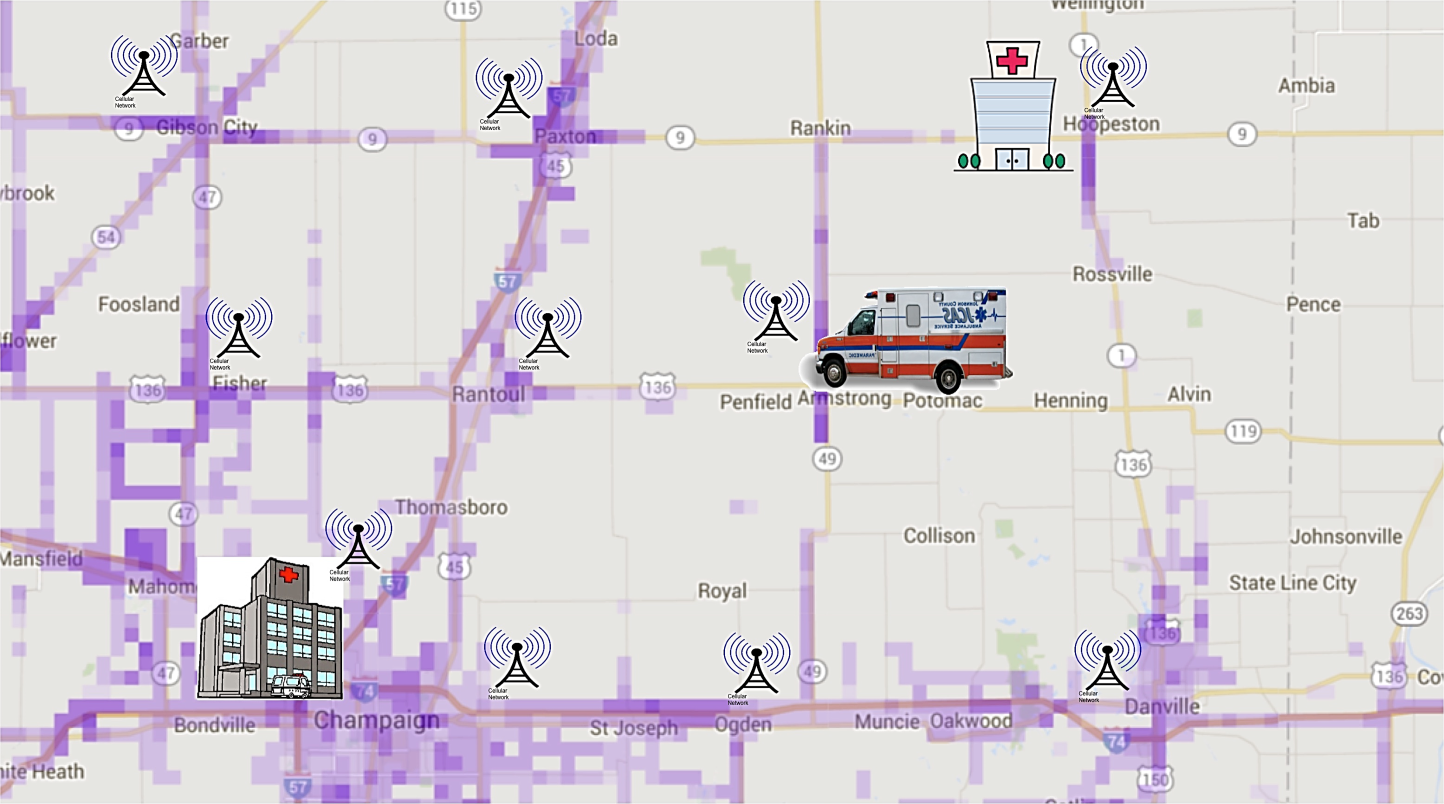}
\caption{Real-world geographical map of our experimental region between Hoopeston rural hospital and Carle center hospital \cite{carle} in Illinois.}

\label{fig:coverage}
\end{figure}

\begin{figure*}[!t]
\centering
\includegraphics[width=.8\textwidth]{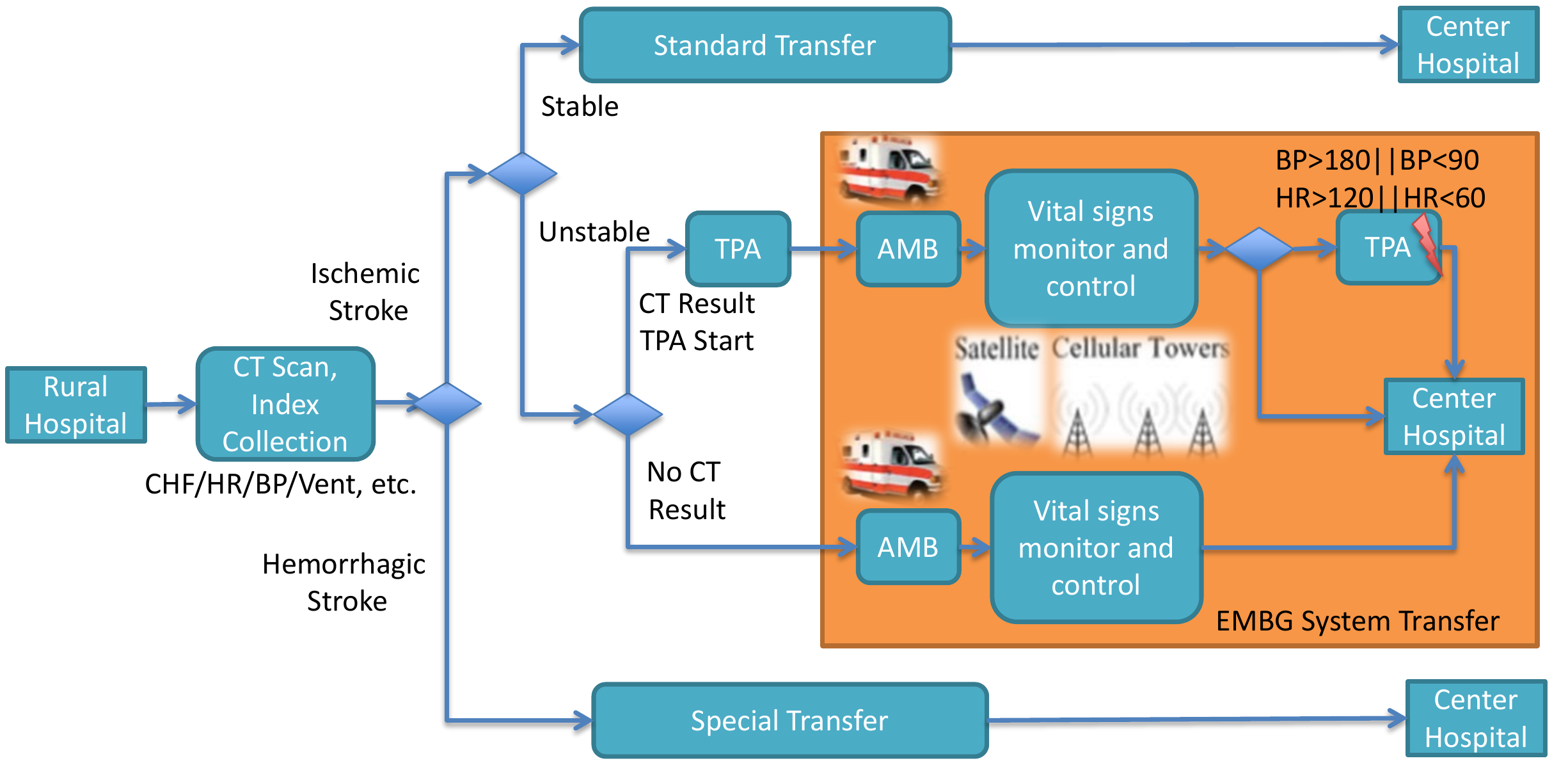}
\caption{Clinical workflow for stroke patient care from a rural hospital to a regional center hospital.}
\label{fig:example-tpa}
\end{figure*}

In this study, on one hand we are motivated by the higher mortality rate of patients in Illinois's rural areas, and on the other hand mainly by the high-fidelity communication requirements of remote patient monitoring and communication breakage in parts of rural routes during high-speed ambulance transport. Figure \ref{fig:coverage} illustrates a real-world geographical map of our experimental region, illustrating 4G mobile data coverage along routes from Hoopeston to Urbana. The maps show varying communication coverage along routes, with some parts suffering communication breakage. From a patient care perspective, transport during critical illness can make route selection patient state dependent. Depending on the type of illness, prompt decisions which weigh a longer more secure bandwidth route versus a shorter, more rapid route with less secure bandwidth must be made. Also, the deadline of the patient transfer is the maximal time that the patient's condition can be kept stable in the ambulance. Since the emergency treatment supervised by remote regional center hospital may effectively extend the transfer deadline, the trade-off between proper route selection and the quality of wireless communication along transport routes becomes an important optimization problem which unfortunately have been neglected by prior work. 

In this paper, we propose a novel physiology-aware route scheduling approach for emergency ambulance transport of patients, especially those in rural areas with acute, high risk diseases in need of continuous remote monitoring. Our physiology-aware route scheduling algorithms aims to select the most optimum route for an en-route ambulance depending on the underlying disease and the criticality of real-time continuous monitoring of patients. We mathematically model the problem into an NP-hard graph theory problem, and approximate a solution based on a trade-off between communication coverage and shortest path in a road network graph. We profile communication along two major routes in a large rural hospital settings in Illinois, and use the collected traces to validate our concept. We preliminary develop our algorithms and evaluate the performance at scale. Our approach is particularly useful to enhance the remote monitoring of patients during high-speed ambulance transport in large rural hospital settings. Our methodology is particularly useful for research groups intersecting with mobile and wireless communications and tele-medicine sub-divisions, and has real-world applications in ambulance dispatch centers especially those targeting emergencies in rural environment. We are planning to validate our adaptive physiology-aware route scheduling system that will serve at central and southern Illinois with 1.2 million people.

The paper is organized as follows. In Section II, we cover some related work and discuss a real clinical use-case for disease-aware ambulance routing. Section III explains our methodology and algorithms. In Section IV, we present our preliminary evaluation experiments, while we conclude the paper in Section V.

\begin{figure*}[!t]
\centering
\includegraphics[width=.9\textwidth]{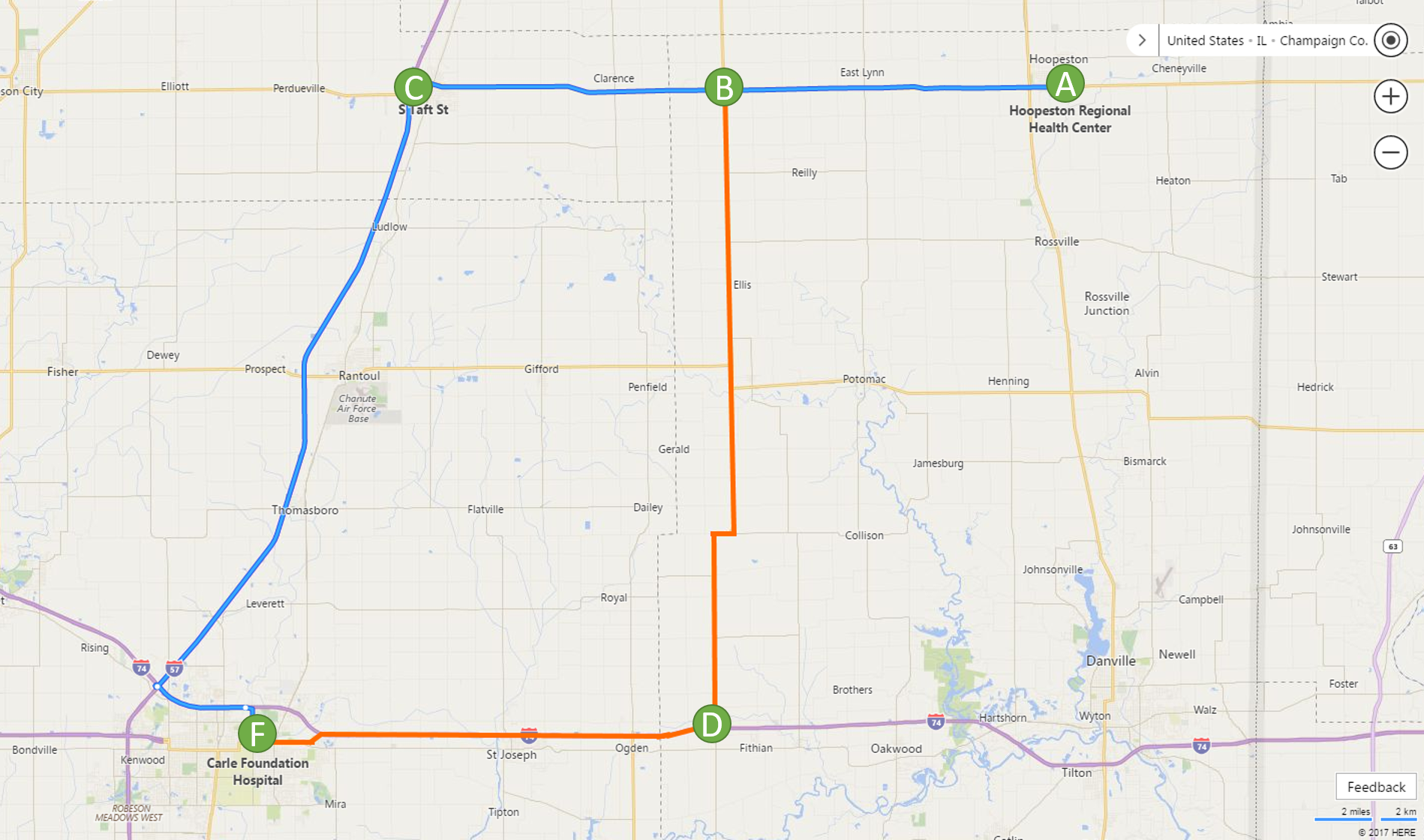}
\caption{Abstract road map of our profiled routes.}
\label{figure:map}
\end{figure*}

\section{Background}
\subsection{Related Work}
There has been a body of work addressing problem of ambulance and medical facility allocation and scheduling. In \cite{amb1} for example, the authors address the problem of allocating ambulances to stations which are distributed throughout a geographical region, aimed to maximize the expected coverage in overall. The core of their decision method is basically crew-shift assignment to allocate crews to shift, based on the maximum number of work hours that can be managed. Similarly, the authors in \cite{amb2} discuss the problem of locating a fleet of ambulances in a given region so that the overall service time is maximized. They model the service time as the coverage of requests by the available ambulance crews. In \cite{amb3}, the authors considered the problem of ambulance allocation to various stations to provide coverage for a given region, assuming that the travel times from a station to the demand location is followed by a specific distribution. In \cite{amb4}, the problem of forecasting the ambulance requests and distributing their working hours is being defined. The authors propose a systematic optimization approach for ambulance scheduling, and discuss how their method can lead to savings and improvement of technicians' schedules. In a similar domain, a wide range of literature review such as \cite{coverage1}, \cite{coverage2}, \cite{coverage3}, \cite{coverage4}, and \cite{coverage5} defined the concept of coverage for emergency facilities, and discussed how it relates to service time, requests, location and re-location in various ways. \cite{coverage6} and \cite{coverage7} provide a literature review on coverage and optimization models for emergency facility location and planning. However, while interesting, none of previous work discuss the problem of emergency ambulance routing in correspondence with communication QoS. Furthermore, previous work mainly discuss a general view of the problem from allocation and optimization, and do not consider the context of medicine. Unfortunately, a relation between routing and disease awareness especially in regards to real-time continuous monitoring has been left unaddressed. Our work in fact complements the previous work by filling the details by proposing scheduling and optimization techniques when a particular service demand (e.g. emergency patient transport) is specifically assigned to a single ambulance.

Due to the recent popularity of vehicular ad hoc networks (VANET), another set of works such as \cite{vanet1}, \cite{vanet2}, and \cite{vanet3} to enumerate a few, focused on the applications of VANET in vehicle routing. The focus of these work is mainly proposing a location-based routing protocol to send information from a source to a destination by performing road selection from a junction to the neighbor junction. These approaches however, not only are general to all moving vehicles and do not consider the emergency requirements of remote care and ambulance patient transfer, but also they focus on position and traffic for data forwarding as opposed to communication coverage for the purpose of road selection. Furthermore, the formulations in these work are based on static topology and employment of offline processes, and therefore topology changes and dynamism are not introduced. Bhoi and Khilar in their work \cite{vanet4} developed a system called VehiHealth used for vehicular communication to transfer data from a source to destination. While their target vehicle is an ambulance and their goal is to provide quick emergency consultancy to save a patient's life, their approach is focused on information and data forwarding which differs from our work on route selection and scheduling. 

\subsection{Real-World Medical Use-cases}
Let's elaborate on emergency care for an acute disease as a real-world use-case and illustrate how ambulance routing can get crucial within the context of remote monitoring. To clarify the concepts, we take acute stroke care being practiced at Carle's hospital networks \cite{carle} as a real-world example of emergency rural ambulance transport.

Stroke is the third leading cause of death and the first leading cause of disability in the United States \cite{stroke-stats, hosseini2016pathophysiological}. In addition, stroke patients are often elderly (in fact, 65\% to 72\% stroke patients are over age 65 \cite{elderly-stroke-rate}) who may need the highest communication requirement for remote monitoring due to complicating medical factors. Furthermore, some effective stroke treatment medications have strict remote monitoring implementation guidelines \cite{prehospitalResearch, oxygenationMonitoring1}; these may begin at the remote facility and continue through ambulance transport to the receiving regional hospital center. Overall, real-time and continuous remote monitoring may take priority depending on type and state of stroke.

Figure \ref{fig:example-tpa} illustrates the workflow for a stroke patient being transferred from a rural facility to a regional hospital center. Let's consider a patient arrives at a rural hospital and the diagnosis of acute stroke is considered. A CT (Computerized Tomography) head scan is performed. The CT images as well as patient's initial neurological examination, laboratory data, and vital signs including heart rate (HR), blood pressure (BP), and oxygenation level are sent electronically to the regional center for interpretation. The diagnosis of an acute stroke is then made, and patient is placed in the ambulance for transport to the regional center hospital.

For the first scenario, with consultation with regional hospital it is determined that the patient has a hemorrhagic stroke (bleeding into the brain from blood vessel rupture), initially manageable in the ambulance. There is noted once in the ambulance a slight deterioration in patient state. In this case, time of transport gets most important. The regional center has the neurological team ready on call to take the patient to radiology, to try to identify the aneurysm and place a stent in the artery to control the hemorrhage. As a result, real-time continuous monitoring gets second priority and ambulance must take the shorter route for fastest transport.

For another scenario, after communication with experts at the remote regional hospital and coagulation lab studies, the patient shows signs of an ischemic stroke. There is no medical contradictions, TPA\footnote{Tissue Plasminogen Activator is a medication to help dissolve blood clots.} is begun and continued onboard the ambulance, and the HR, BP, oxygenation, and neurological status are remotely monitored. With ischemic stroke, highest priority is maintaining the patient's vital signs. The HR and BP in specific must be kept within strict limits. The BP assumes particular importance if it rises too high (greater than 180/-) or falls too low (less than 90/-). Vigorous real-time monitoring o continuous intravenous infusion of active medications to lower blood pressure using nicardipine or nitroprusside medications is required during transport. In this case, time is not critical, and in the face of stable patient and vital signs, the network connection becomes most critical. Precise monitoring and management of BP, TPA, and oxygenation (O2 saturation) is necessary if brain condition deteriorates \cite{oxygenationMonitoring1, oxygenationMonitoring2}. As a result, the ambulance may take the longer route, but more secure with coverage to assist EMT manage change in patient state in accordance with guidance from remote physicians.

\begin{figure*}[!t]
\centering
\includegraphics[width=1\textwidth, trim = 255 80 300 80, clip = true]{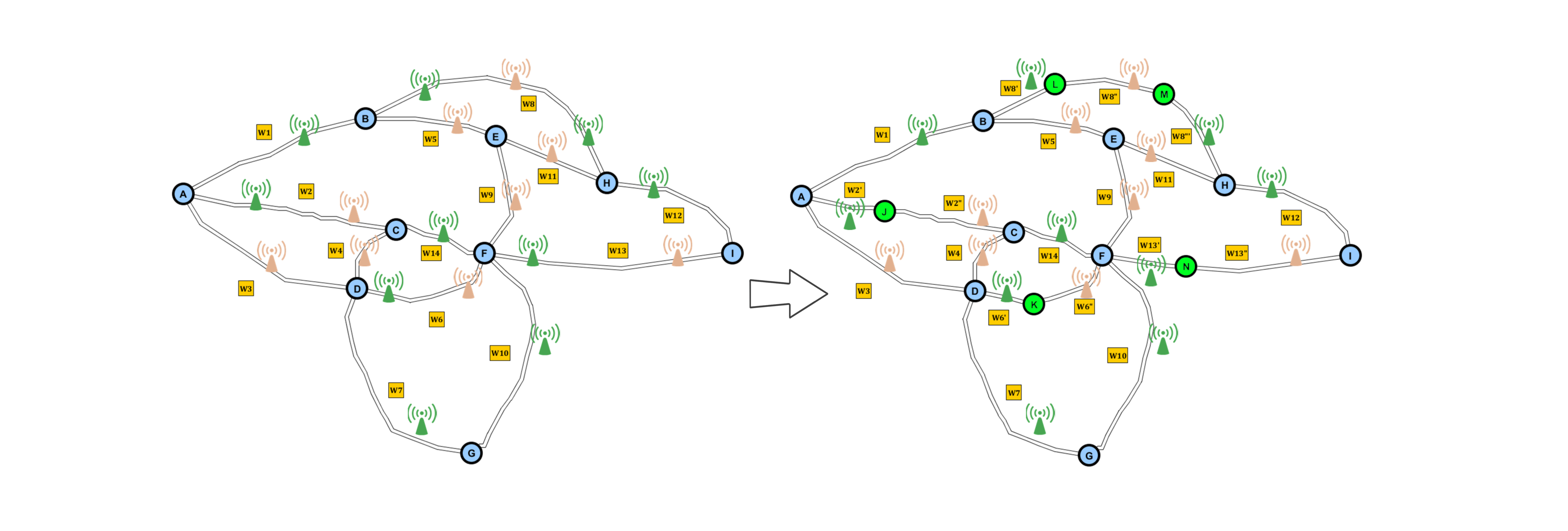}
\caption{An example planar graph model. (Left) The initial abstract graph with partial communication on edges. (Right) The transformed graph model with added nodes (highlighted) and each edge having a single communication label.}
\label{figure:graphs}
\end{figure*}

Figure \ref{figure:map} illustrates the geographical trajectories of the two major routes from Hoopeston's rural hospital (point A) to Carle regional hospital in Urbana (point F). Route $A\to B\to C\to F$ illustrates a longer route, but with higher-fidelity communication and more secure coverage according to Figure \ref{fig:coverage} and our conducted profiling. This longer route is most suitable for acute diseases with high real-time monitoring requirements such as those seen in ischemic stroke patients. In contrast, while the transport duration is shorter in the route $A\to B\to D\to F$, the communication coverage bears much lower fidelity, some parts with no coverage, which makes it most suitable for transport of patients with hemorrhagic stroke. Overall, a continuous and high-quality communication coverage gets crucial depending on type of illness and therefore, how critical remote monitoring is. This requires the ambulance to adaptively select best transport route in response to changes in patients' physiological states and communication coverage conditions. 

\section{Methodology}
\subsection{Problem Definition}
For continuous monitoring of patient when an ambulance is en route, an important feature is reliable high-speed communication coverage along transport routes. The design of our physiology-aware route scheduler is centered around communication coverage along transport routes to assist ambulances pick the best possible route when communication is needed for real-time remote monitoring of patients. The selection of a shorter route with low-fidelity communication coverage versus a longer route with better coverage and thus, better remote monitoring and assistance is what we call a communication-aware shortest path problem.

Our route selection problem is generally considered a resource-constrained shortest path in the sense that it seeks a shortest and fastest path in a directed graph with pre-defined edge lengths from an origin node to a destination node, but with regards to resource limitations. The edge lengths are associated with ambulance travel duration, and the route selection is subject to one or more constraints. To illustrate the concept of resource constraints better, a constraint can be that the total amount of some data that has to be collected at the nodes along the paths be less than a given limit. In a similar way, in our communication-aware shortest path problem, an ambulance needs to pick a path of minimum length from any given origin node $s$ to a given destination $t$ to certain constraints. For an ambulance en route from a rural hospital to a city hospital, we define a set of requirements for a path constraints as follows:

\begin{itemize}
\item \textbf{Req. 1:} The total length of communication unavailability and breakage in a path must not exceed a specific duration constraint $D_1$.
\item \textbf{Req. 2:} the total length of \textit{continuous} communication unavailability and breakage in a path must not exceed a specific duration constraint $D_2$. 
\end{itemize}

Our communication-aware shortest path problem therefore seeks a trade-off between the shortest and fastest transport time and satisfaction of communication constraints as mentioned above. Similar to multi-criteria decision problems, shortest path problems with resource constraints are NP-hard in the strong sense. However, efficient approximation algorithms can be utilized, so this approach is computationally feasible. 

The standard approach to solve shortest path problems with resource constraints is based on dynamic programming. Our communication-aware shortest path problem however, differs from the known resource-constraints shortest path problems. By definition, a resource is constrained if there is at least one \textit{node} in the graph where the resource must not take all possible values. However, our problem modeling has multiple variations and unique features as we describe in the following:
\begin{enumerate}
\item \textit{Edge-constrained shortest-path problem:} In our context, the notion of resource is not applied to the node, but is rather applied to the edges. This variation complicates the problem and makes it non-deterministic as the window of resources or the set of allowed values for the resource is defined at each traversed edge. This implies that a prior insight of the future to-be-chosen edge is needed before selecting an edge. Therefore, if edge $(i,j)$ from node $i$ to node $j$ is chosen, then selecting the next edge $(j,k)$ from node $j$ to node $k$ is considered valid \textit{only if} the whole resulting path from node $i$ to node $k$ is valid and falls within the communication requirements. In other words, assuming $Fn(i,j)$ to be a function returning the optimum route from a given node $i$ to $j$, then in our context, $Fn(i,j) \neq Fn(i, j-1)+ Fn(j-1, j)$, which differs from a dynamic programming approach.

\item \textit{Partial communication on edges:} In our context, communication coverage can exist partially on a route or an edge. This also complicates our problem since multiple values are associated with a single route as opposed to a single value for an individual route. Figure \ref{figure:graphs} (left) illustrates an example road network planar graph, with routes having a weight label representing the transport duration as well a binary label representing the communication coverage that can cover parts of a route. A green coverage symbol represents availability of coverage whereas a red symbol showing lack of coverage on a part of an edge. For instance, in Figure \ref{figure:graphs} (left), edge $e_{AC}$ partially carries two communication coverage labels, with a portion with communication coverage and the rest with lack of communication coverage.

\item \textit{Mutually adaptive route scheduling:} Our problem modeling is more complex due to the adaptation nature of the routing algorithm. The algorithm must be adaptive in the sense that once a node is reached, the pre-determined optimum path can be modified depending on current conditions. This makes the routing algorithm interactive and mutually adaptive with physicians at the remote center hospital.

\item \textit{Trade-offs in satisfaction of requirements:} Due to the trade-off nature of our problem modeling, satisfaction of requirements must be relative to the transport duration. In other words, satisfying a specific requirement, for instance Req. 1, must not enforce the algorithm to select an extended route with a long duration. For that purpose, we devise an objective function as a metric to weigh alternative paths for the purpose of optimization. In the graph represented in Figure \ref{figure:graphs}, there are three alternative paths from $D$ to $F$. While there is communication breakage in parts of paths $(D\rightarrow C\rightarrow F)$ and $(D\rightarrow F)$, they are shorter than the third alternative path $(D\rightarrow G\rightarrow F)$ which has full communication coverage. Therefore, our objective function must provide a metric to weigh these three paths differently.
\end{enumerate}

\subsection{Proposed Solution}
The underlying system of transport routes is characterized by a weighted directed graph $G = (\mathcal{T}, E)$ representing a graph of $|\mathcal{T}|$ nodes and $|E|$ links between them. Thus, each vertex $\tau_i \in \mathcal{T}$ represents an intersection, and each edge $e_{ij} = (\tau_i, \tau_j) \in E$ represents a possible route between a pair of vertices, $\tau_i$ and $\tau_j$. Due to the fact that communication coverage might exist for parts of an edge, we use labels to store information on the communication coverage for partial paths. Aside from having the length of an edge, a label considers the level of communication coverage along that edge, which maps the set of all possible values of communication bandwidth to a set of limited integer values defined as a resource. In this initial work, we start with a binary label for communication, specifying the availability of communication coverage.

In our generalized route selection setup, each edge $e_{ij}$ is associated with a weight $w_i$ representing the length or duration of transport time, and multiple labels each denoted as a binary flag $v_i \in \{0,1\}$ representing existence or lack of communication coverage. The values of $w_i$ and $v_i$ are known in advance based on profiled communication traces. The problem of selecting the optimal set of upcoming route in order to select the most proper shortest path based on communication needs can be formulated as an optimization problem. The goal is to minimize a cost function derived by transport duration as well the sets of requirements as defined previously through selectively choosing a particular subset of routes, $S \subseteq E$, under which the total cost is minimized.

First, to accommodate the partial communication coverage on edges with possibility of edges having multiple communication labels, we transform $G$ into the planar graph $G'$ through an edge partitioning process. All multi-labeled edges are partitioned into multiple independent edges, each with a single communication label, through inserting additional nodes between each pair of edge segments.

Figure \ref{figure:graphs} (right) shows the transformed graph of the example planar graph model shown in Figure \ref{figure:graphs} (left), with added vertices and extra edges, with each edge associated with a single coverage label. It illustrates how the edge $e_{DF}$ with weight $w6$ and two communication labels is segmented into two distinct edges, $e_{DK}$ with weight $w6'$ and $e_{KF}$ with weight $W6''$, each carrying separate communication labels. The new added nodes are highlighted in green. 

Next, to keep the number of candidate paths as small as possible, we perform a graph simplification approach. To achieve that, we iterate through all edges $e_{ij} \in E$ and remove edges violating Req. 2. Due to the criticality of this requirement, Req. 2 is checked as a satisfaction rule to filter candidate paths.

In the next step, an objective function is defined as a metric to weigh various alternative paths. Our objective function seeks trade-offs between a path which is long, but with less connectivity, and a shorter, but more secure connectivity. Therefore, we define our objective function as a weighted function of both the total length and total communication coverage. In this pilot study, we start from the simplest function, and define the objective function $f(p)$ for a given path $p$ as the sum of the total length of the path and the total length of breakage multiplied by a specific co-efficient, represented as:
\begin{align}
\label{eq:objective}
f(p)=\left(\sum W_{(i,j)} + \alpha . \sum W_{(m,n)}\right)\\
~~s.t.~~V_{(m,n)}=1,~(i,j),(m,n) \in p.
\end{align}
$\alpha$ a disease-specific coefficient given by physicians which determines how critical continuous remote monitoring is for a specific illness at any given time. 

\subsection{Algorithms}

\begin{figure*}[!t]
\centering

\includegraphics[width=1.02\columnwidth, trim = 50 260 50 265, clip = true]{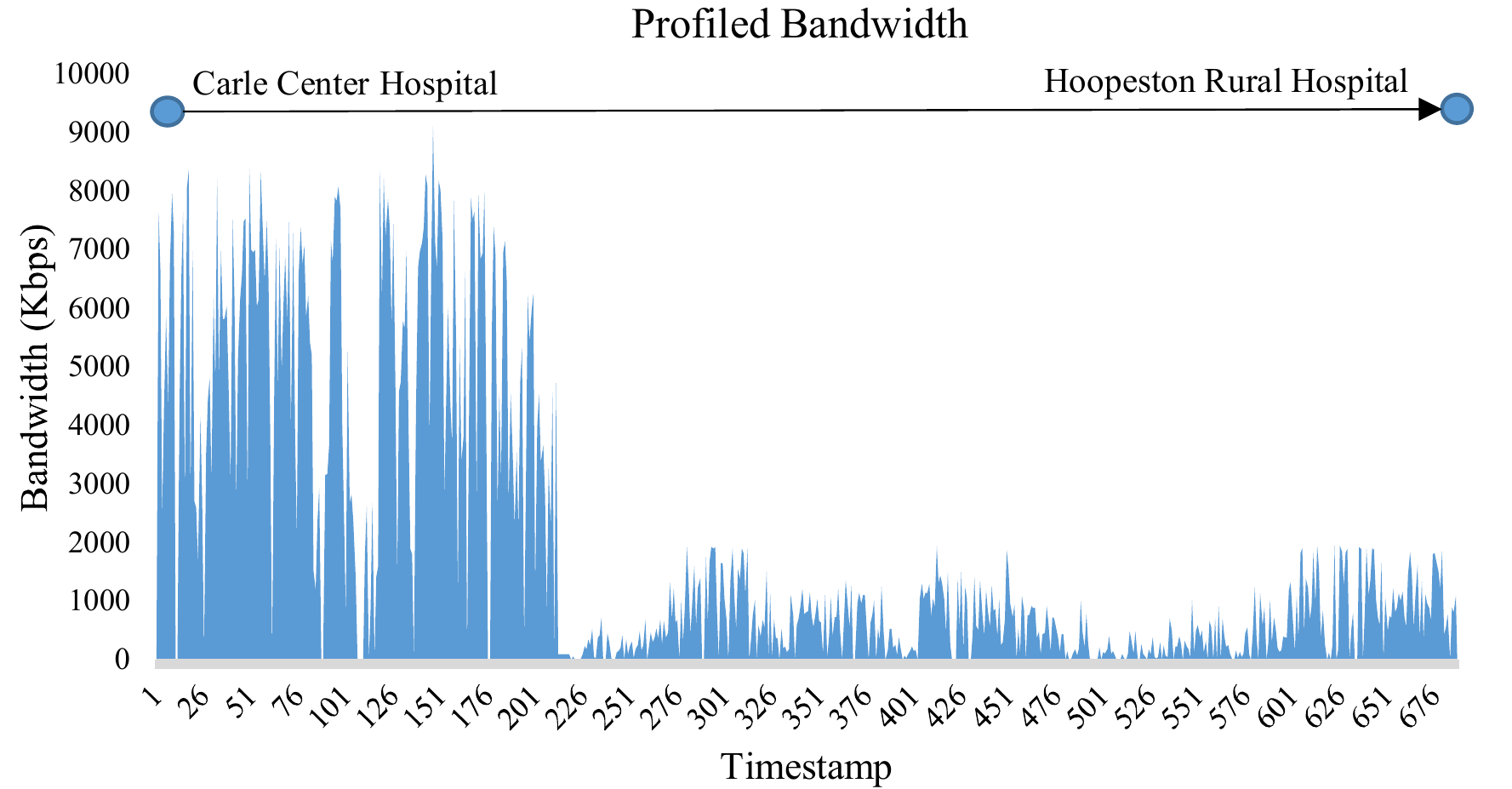}\includegraphics[width=1.02\columnwidth, trim = 50 260 50 265, clip = true]{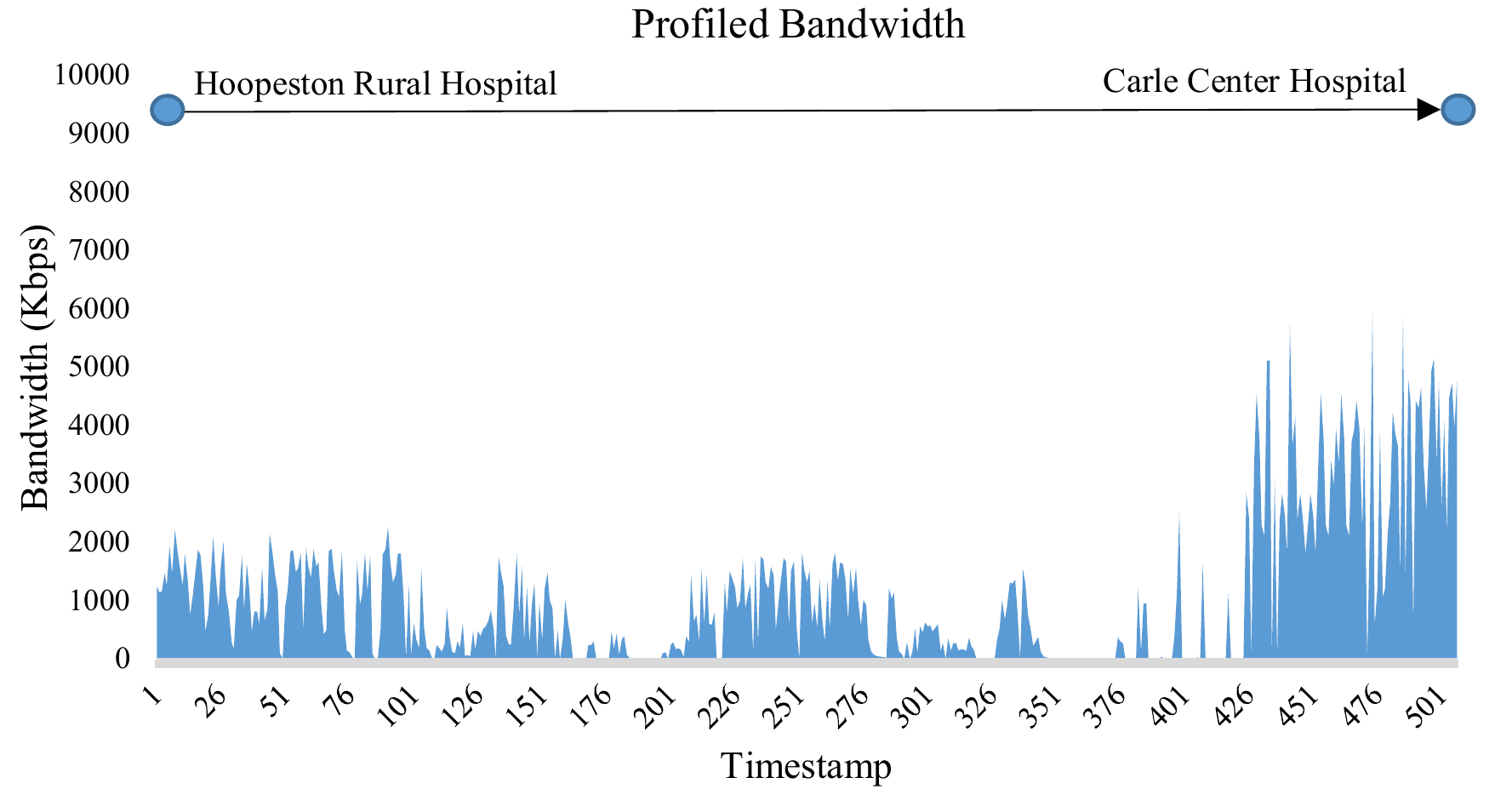}
\caption{A sample of our profiled communication from Hoopeston rural hospital to Carle center hospital.}

\label{profile}
\end{figure*}

Our proposed solution involves finding candidate shortest paths between a designated pair of nodes in a given directed weighted graph. For that purpose, use Eppstein's K-shortest path algorithm \cite{eppstein} to find $k$ number of shortest paths as it represents all possible deviations from the shortest path. The core of the proposed approach is to use the K-shortest path algorithm to shortlist candidate paths, and apply the objective function as a metric to select most suitable path. If no candidate paths are found given the requirements, the algorithm is repeated with higher threshold values until a path is returned. Overall, our proposed algorithm is consisted of five major phases, as follows:
\begin{itemize}
\item \textbf{Phase 1.} Graph transformation phase. Transform the initial multi-labeled graph $G$ into a singly-labeled graph $G'$ through the edge partitioning process as described previously and depicted in Figure \ref{figure:graphs}.

\item \textbf{Phase 2.} Graph simplification phase. The transformed graph $G'$ is then simplified, and transformed into graph $G''$ through ignoring edges violating Req. 2.

\item \textbf{Phase 3.} K-shortest paths candidates. Run our K-shortest path algorithm, and modify the Eppstein's K-shortest path algorithm to retrieve each shortest path one-by-one, and verify the path against Req. 1 and Req. 2 until we get $k$ distinct shortest paths from the designated source to the destination node. Finally the total lengths of each path is calculated and stored.
\begin{itemize}
\item If no path is found, increment $D_1$ by a coefficient $d$ with an upper bound $D'$, so that $D_1=d.D_1 ~s.t.~ D_1\leq D'$. Repeat until $k$ paths are found. 
\item If no path is found, the problem can not be solved. The EMT must communicate with the physicians for an acceptable solution. This can be following the best effort such as shortest path.
\end{itemize}

\item \textbf{Phase 4.} Repeat phase 2 for $h$ discrete values of $D_{1i} \in [\beta_1. D_1, \beta_2.D_1]: 1\leq i \leq h$ and construct a 2-dimensional array, with each row representing an array of the resulting k-shortest paths for a given $D_{1i}: 1\leq i \leq h$.
\begin{itemize}
\item Note that this algorithm retrieves $k$ paths not violate Req. 2.
\end{itemize}

\item \textbf{Phase 5.} Lastly, apply our objective function in Equation \ref{eq:objective} against each individual path in the constructed 2-dimensional array and calculate the values. Select the path resulting in the lowest cost as the optimal path.
\end{itemize}
It should be noted that the values of $k$, $d$, $\beta_1$ and $\beta_2$ are defined by the physicians and can be adjusted based on the criticality of a disease as well as the complexity of a given planar graph.

\section{Evaluation}

To validate our study in real world and to facilitate a logical interpretation of the problem to be analyzed, we conducted communication coverage and bandwidth profiling in a large region of hospital health system in Illinois. We took a scenario where a patient is transferred from a rural hospital to a regional center hospital via a high-speed ambulance, and profiled the two major routes in Figure \ref{figure:map}, from Hoopeston's rural hospital to Carle Foundation Hospital in Urabana as proof of concept. To profile the available bandwidth, we developed a mobile bandwidth profiler application in collaboration with Carle Ambulance Service, and collected communication bandwidth traces under 4 major US cellular networks \cite{hosseinidataset}.

Figure \ref{profile} demonstrate only a small subset of our profiled data under Sprint cellular network. Figure \ref{profile} (Left) shows the downlink bandwidth for the longer route (route $A\to B\to C\to F$) shown in Figure \ref{figure:map}, whereas Figure \ref{profile} (Right) shows the bandwidth traces for the shorter route (route $A\to B\to D\to F$) shown in Figure \ref{figure:map}. The vertical axis shows the available bandwidth while the horizontal axis shows the timestamp with each point of data accounting for the four seconds of sampling period. As can be witnessed, interestingly while the route $A\to B\to D\to F$ is shorter with less transport duration (almost 35 minutes), it involves more vigorous communication breakage, making it more suitable for emergency scenarios where transport duration is of higher criticality than the remote monitoring. On the contrary, the longer route  $A\to B\to C\to F$ (almost 47 minutes) shows a more reliable and more continuous communication coverage in spite of longer transport duration, which makes it more suitable for emergency ambulance transports where remote monitoring becomes more critical.

\begin{figure}[!t]
\centering
\includegraphics[width=.95\columnwidth, trim = 50 260 50 260, clip = true]{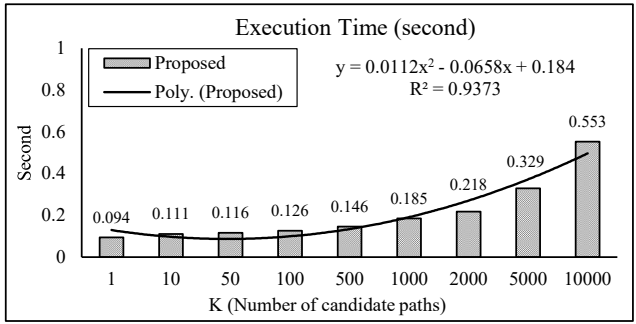}
\caption{Scalability analysis of proposed algorithm in terms of execution time.}

\label{time}
\end{figure}

We evaluated the execution time of our algorithm, and conducted preliminary computational experiments to analyze the average-case performance and scalability of the algorithm. Our experimental platform was a HP Z230 SFF Intel Xeon E3-1240 machine with 3.4 GHz quad-core processors, 8 MByte Cache and 8,192 MByte physical memory running Java 1.8.0.121. We randomly generated a large-scale planar directed weighted graph representing a road network, with around 12,000 nodes and 30,000 edges with non-integer weights. We generated random source and destination nodes, and repeated each experiment 10 times and calculated the average execution times.

Figure \ref{time} shows a subset of our experimental evaluation. It illustrates the average execution times for different sizes of $k$ (i.e. number of shortest paths analyzed), varying from 1 to 10,000, with $d$ and $\beta$s set to 1 for the purpose of scalability analysis. This means the biggest problem solved was analyzing the top 10,000 shortest path. The curve line shows the polynomial trend-line for the execution time, indicating that with a probability of higher than 0.93, the function $y=0.0112x^2 - 0.0658x + 0.184$ can generalize the overall trend and estimate future data points. The results illustrate the efficiency of our proposed algorithm at scale, and verifies its suitability for the real-time requirements of large-scale route scheduling. We will comprehensively present all our evaluation results and profiled traces in future studies.

\section{Conclusions and Future Work}
Use of telecommunication technologies for remote monitoring of patients can enhance effectiveness of emergency ambulance transport from rural areas to a regional center hospital. However, the communication along the roads in rural areas can range from 4G to 2G with some parts with lack of communication coverage, which poses a major challenge in remote supervision of patients in ambulances. Depending on the type of disease, rapid decisions which weigh a longer more secure bandwidth route versus a shorter, more rapid route with less secure bandwidth must be made. Unfortunately, the trade-offs between route scheduling and the quality of wireless communications in the context of disease becomes an optimization problem which unfortunately have been neglected in previous studies.

In this paper, we propose a physiology-aware route scheduling algorithm for emergency ambulance transport of patients with acute diseases in need of continuous remote monitoring. We study the trade-offs between route scheduling and the quality of wireless communications in the context of disease awareness and investigate it through an NP-Hard optimization problem. We design our algorithms, validate them conceptually using our real-world profiled data, and run preliminary experiments to evaluate the runtime performance and scalability. We believe that our techniques can enable next-generation ambulance dispatch centers in emergency patient transfer scenarios when continuous network coverage is critically needed, which can help drop morbidity and mortality rates with early diagnosis and effective treatment.

In the future, we plan to explore the effects of variability of the parameter $\alpha$ in Eq. 1 on the algorithm, and study how results change (as well as processing time) according to the arbitrary choices of $\alpha$ values.

\section{Acknowledgments}
This research is supported in part by NSF CNS 1329886, NSF CNS 1545002, and ONR N00014-14-1-0717.

\bibliographystyle{IEEEtran}
\bibliography{sigproc}

\end{document}